\documentclass[12pt]{article}
\usepackage{amsmath}
\usepackage{amssymb}
\usepackage{cite}

\usepackage{epsfig}

\psfull

\newcommand{\ie}{\emph{i.e.}\ }

\def\order#1{{\cal{O}}\left(#1\right)}

\def\HR{{\cal H}_{\mathrm{R}}}
\def\HL{{\cal H}_{\mathrm{L}}}

\def\qbar{{\bar q}}
\newcommand{\Ft}{{\widetilde F}}
\newcommand{\Pt}{{\widetilde P}}

\newcommand{\heavy}{\mathrm{heavy}}

\newcommand{\rightH}{\mathrm{right}}

\def\cS{{\cal{S}}}    
\def\cC{{\cal{C}}}    
\def\cD{{\cal{D}}}    

\newcommand{\tot}{\mathrm{tot}}

\def\CF{C_F}

\def\CA{C_A}
\def\NC{N_C}
\def\nf{n_{\!f}}

\def\as{\alpha_{{\textsc{s}}}}

\def\asb{{\bar \alpha}_{{\textsc{s}}}}

\def\ee{e^+e^-}

\begin{document}
\begin{titlepage}
\begin{flushright}
{DESY--01--055 \\
  CERN--TH/2001--110\\
  LPTHE--01--21\\
  hep-ph/0104277 \\
  April 2001
}
\end{flushright}              
\vspace*{\fill}
\begin{center}
{\Large 
\textsf{\textbf{Resummation of non-global QCD observables.}}\footnote{Research
    supported in part by the EU Fourth Framework 
    Programme `Training and Mobility of Researchers', Network `Quantum
    Chromodynamics and the Deep Structure of Elementary Particles',
    contract FMRX-CT98-0194 (DG 12-MIHT).}}
\end{center}
\par \vskip 5mm
\begin{center}
       
       {\large \textsf{ M.~Dasgupta}} \\ 
          DESY, Theory Group, Notkestrasse 85, Hamburg, Germany.
          \vspace{0.5cm}\\
       {\large \textsf{  G.~P.~Salam}} \\
  CERN, TH Division, 1211 Geneva 23, Switzerland.\\
  LPTHE, Universit\'es P. \& M. Curie (Paris VI) et Denis Diderot
  (Paris VII), Paris, France.
 
\end{center}
\par \vskip 2mm
\begin{center} {\large \textsf{\textbf{Abstract}}} \end{center}
\begin{quote}
We discuss issues related to the resummation of non-global
  observables in QCD, those that are sensitive to radiation in only a
  part of phase space. Examples of such observables are certain
  single-hemisphere event shapes in $\ee$ and DIS. Compared to global
  observables (those sensitive to all emissions, e.g.\ the $e^{+}e^{-}$ thrust)  a
  new class of single-logarithmic terms arises. These have been
  neglected in recent calculations in the literature. For a whole set
  of single hemisphere $\ee$ and DIS event shapes, we analytically
  evaluate the first such term, at order $\as^2$, and give numerical
  results for the resummation of these terms in the large-$\NC$
  limit.
\end{quote}
\vspace*{\fill}

\end{titlepage}


\section{Introduction}

It has long been known that in certain restricted regions of phase
space the coefficients of the QCD perturbation expansion can be
parametrically large. For semi-inclusive observables this is often due
to the incomplete cancellation of soft and/or collinear logarithms.

A set of variables particularly sensitive to this kind of problem is
event-shape distributions $\frac{1}{\sigma} \frac{d\sigma}{dv}$ 
which we shall use as an illustration throughout.
Typically one finds that the
perturbative series is dominated by terms $(\as^n \ln^{2n-1} v)/v$,
where $v\ll1$ denotes the value of the event-shape variable. In order
to have a meaningful perturbative description these terms have to be
resummed to all orders. This has been done for a range of observables
\cite{CTTW,CTW,JetRates,Cpar,NewBroad,DIStauz,rhoLight,KoutLong,KoutLett}.

Most of the observables considered so far have the property of
exponentiation, implying the following structure for the resummation,
\begin{equation}
  \label{eq:basicresummed}
  \Sigma(v) = 
  \left(1 + \sum_n C_n \asb^n\right) e^{Lg_1(\as L) + g_2(\as
      L) + \cdots} 
     + D(v),
\end{equation}
where the distribution is given by $d\Sigma/dv$.  The remainder
function $D(v)$ has the property that it goes to zero for $v\to0$;
$L = \ln 1/v$ and $\asb = \as/2\pi$. 
The terms resummed by $L g_1(\as L)$ are referred to as
leading (or double) logarithms (LL), while those in $g_2(\as L)$ are
next-to-leading (or single) logarithms (NLL). For certain variables,
rather than a single exponent one finds a sum or integral over exponents
\cite{KoutLong,KoutLett,rhoLight}.

A property common to most of the observables considered until recently
is that they are sensitive to emissions anywhere in phase space. We
call these \emph{global} event shapes. Examples are the thrust, the
heavy-jet mass, the $C$-parameter. In this paper we are interested in
\emph{non-global} observables, those sensitive to radiation in only a
part of phase space. We shall discuss mainly single-hemisphere
event-shapes in $\ee$ (especially the jet-mass). Other examples
include a number of DIS current-hemisphere observables and quantities
such as the mass and broadening of single jets in multi-jet ensembles.

The literature on the subject of global event shapes often makes use
of a combination of results in separate hemispheres to obtain the
properties of the whole event \cite{CTTW,CTW,NewBroad}.  When the
separate hemisphere results are combined in this fashion the answer
obtained is correct. However this can be misleading, since it turns
out that the existing single-hemisphere results, applied literally to
just a single hemisphere event-shape, are incomplete at NLL level.
This has not been taken into account in the recently presented results
for the light-jet mass and the narrow-jet broadening \cite{rhoLight}
and for the narrow-jet thrust-minor \cite{KoutLong}. In the case of
the first two variables the incompleteness of the results can be seen
from a comparison of the $\order{\as^2 L^2}$ term of the expansion of
the resummation with fixed-order predictions from a Monte Carlo
program such as Event2 \cite{CSDipole}.\footnote{Here we refer to the
  original version of \cite{rhoLight} where it was stated that terms
  down to $\as^n L^{2n-2}$ of the expanded result were under control.
  In a revised version (v2) which appeared on the hep-ph arXiv shortly
  before the submission of our paper, the claims regarding the
  accuracy were revised to state that only terms down to $\as^n
  L^{2n-1}$ are controlled. In our notation that corresponds to
  controlling only the $\as L$ term of $g_2$, but not terms $\as^n
  L^n$ with $n\ge2$, nor $C_1$. Here we follow the convention adopted in
  \cite{CTTW} rather than that in \cite{rhoLight}, and use the
  term next-to-leading to refer to
  logarithms in the exponent(s), not in the expansion of the
  exponent.}

To illustrate the point it is convenient to study a simpler variable,
the hemisphere jet-mass, defined as 
\begin{equation}
  \rho = \frac{\left(\sum_{i \in \mathrm{hemisphere}}
        k_i\right)^2}{\left(\sum_i E_i\right)^2} ,
\end{equation}
where the hemisphere is taken with respect to the thrust axis.  Once
the distribution of $\rho$ has been determined, it can be combined
with the known heavy-jet mass distribution \cite{CTTW} to obtain the
light-jet mass distribution, using the following exact relation
\begin{equation}
  \label{eq:heavylight}
  \frac{d\sigma}{d\rho_\mathrm{light}} +
  \frac{d\sigma}{d\rho_\mathrm{heavy}} = 
  \frac{d\sigma}{d\rho_\mathrm{left}} +
  \frac{d\sigma}{d\rho_\mathrm{right}} \equiv 
  2\frac{d\sigma}{d\rho}\,,
\end{equation}
where $\rho$ is the hemisphere jet-mass.  An analogous relation
applies for the jet-broadenings. For the narrow-jet thrust minor the
situation is more complex but considerations similar to those
addressed here will apply.

\begin{figure}[tb]
  \begin{center}
    \resizebox{\textwidth}{!}{\input{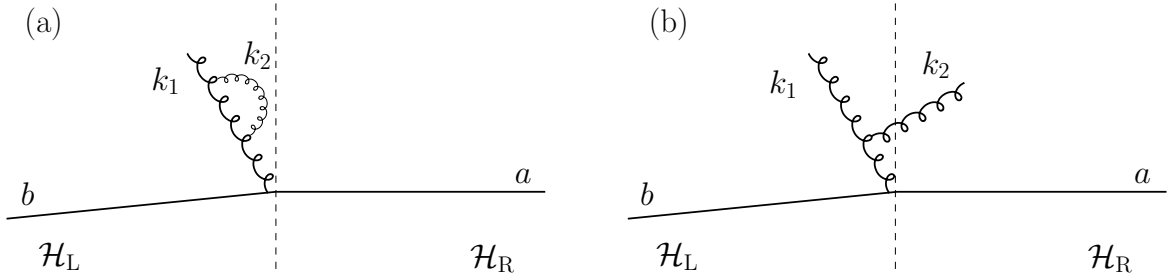}}
    \caption{Kinematic configurations of interest}
    \label{fig:twogluons}
  \end{center}
\end{figure}

One can understand the origin of the qualitative difference between
the heavy-jet and single-hemisphere mass resummations by examining
figure~\ref{fig:twogluons}, which depicts two soft large-angle gluons,
whose energies are ordered $Q \gg k_1 \gg k_2$. For the heavy-jet mass
there is a cancelation between the two contributions because the
emission of $k_2$ does not affect the value of $\rho_\heavy$ (which is
determined by the harder emission, $k_1$). In the case of the
right-jet mass, $k_1$ has no effect on $\rho_\rightH$, so the event
shape receives a contribution only from the softer emission, $k_2$.
This spoils the cancelation with the virtual diagram,
figure~\ref{fig:twogluons}a, thus leading to a term $\as^2 L^2$, i.e.\ 
a NLL effect. Such effects are a general feature of non-global
observables.

At all orders, to NLL accuracy, the form of the resummed jet-mass
distribution will be
\begin{equation}
  \label{eq:finalform}
  \Sigma(\rho) = \left(1 + \asb C_1^{(q)} \right) \cS(\as L)\,
  \Sigma_q(\as, L) + \asb C_1^{(g)}  \Sigma_g(\as, L) + D(\rho)\,.
\end{equation}
$\Sigma_q$ and $\Sigma_g$ are the resummed jet-mass cross sections
which can be obtained by integrating the quantities $J_q$ and $J_g$
first derived in \cite{CT} and widely used in the literature. They
have structures like that of the exponential factor in
eq.~\eqref{eq:basicresummed}.  The sum of two resummed contributions
arises because in a fraction $\order{\as}$ of events the hard particle
in the right hemisphere ($\HR$) is a gluon.

The
new effect considered in this article is embodied by the
function $\cS$.\footnote{In principle a similar function multiplies
  $\Sigma_g$, but at NLL accuracy it can be neglected.} We write its
expansion as
\begin{equation}
  \cS(\as L) = 1 + \sum_{n = 2} S_n\, (\asb L)^n\,.
\end{equation}
It is straightforward to exactly compute the first non-trivial term
$S_2$ and this is done in the following section. The full computation
of $\cS$ involves considering an ensemble of an arbitrary number of
large-angle energy-ordered soft gluons in $\HL$, which coherently emit
a single, softer gluon into $\HR$.  For reasons elucidated later it is
difficult to carry out an all-orders treatment of such an effect
analytically.  We therefore opt to treat these effects using a Monte
Carlo algorithm valid in the large-$\NC$ limit.  This is outlined in
section~\ref{sec:AllOrders} and further details are given in the appendix.

Finally in section~\ref{sec:Event2} we compare our results to the
$\order{\as^2}$ predictions of Event2. Phenomenological predictions
including this effect will be shown elsewhere \cite{DSDIS}.

\section{Fixed order calculation}
\label{sec:FixedOrder}

First we calculate the contribution to the jet-mass distribution from
the configuration in figure~\ref{fig:twogluons}b, considering the
right-hemisphere jet for concreteness. We introduce the following
particle four-momenta
\begin{subequations}
\begin{align}
  k_a &= \frac{Q}{2}(1,0,0,1)\,, \\
  k_b &=  \frac{Q}{2}(1,0,0,-1)\,, \\
  k_1 &= x_1\frac{Q}{2}(1,0,\sin\theta_1,\cos\theta_1)\,,\\
  k_2 &= x_2\frac{Q}{2} (1,\sin\theta_2 \sin\phi,\sin\theta_2
  \cos\phi,\cos\theta_2)\,,
\end{align}
\end{subequations}
where we have labelled the quark and antiquark as $a$ and $b$ and
defined energy fractions $x_{1,2} \ll 1$ for the two gluons.  We have
ignored recoil in the kinematics, because the jet-mass is insensitive
to it.

When gluon $2$ is in $\HR$ the jet mass has the value $\rho =
x_2(1-\cos\theta_2)/2$. When only the quark is in $\HR$, $\rho=0$.

We write the matrix element for ordered two-gluon emission as (see for
example \cite{DMO})
\begin{subequations}
\begin{align}
  W &= 4\CF \frac{(ab)}{(a1)(1b)} \left( \frac{\CA}{2}
    \frac{(a1)}{(a2)(21)} + \frac{\CA}{2} \frac{(b1)}{(b2)(21)} +
    \left(\CF - \frac{\CA}{2}\right)\frac{(ab)}{(a2)(2b)} \right)
  \\
  &= C_F^2 W_1 + C_F C_A W_2\,,
\end{align} 
\end{subequations}
where $(ij) = k_i \cdot k_j$. The result is valid for $1 \gg x_1 \gg x_2$
as well as for the opposite ordering of the gluons, and furthermore is
completely symmetric under interchange of $k_1$ and $k_2$. We have
however chosen to write it in an asymmetric form to emphasise the
dipole structure of the emissions, namely radiation of gluon $k_1$
from the $ab$ dipole, followed by the radiation of gluon $k_2$ from
the $a1$, $1b$ and $ab$ dipoles.

The $\CF^2$ piece of the matrix element, $W_1$ corresponds to
independent gluon emission and is included in the usual resummation of
the quark jet-mass, \ie in $\Sigma_q$ of eq.~\eqref{eq:finalform}. To
study specifically the modification relative to the standard quark-jet
mass result due to configurations like that shown in
figure~\ref{fig:twogluons}b, one must consider the $C_F C_A$
part, $W_2$, of the emission probability above.

The integral to be considered, related to the probability for
the jet mass to be less than $\rho$, is then
\begin{multline}
  -C_F C_A\left ( \frac{\alpha_s}{2 \pi} \right )^2\int_{-1}^0
  d\!\cos{\theta_1} \int_{0}^{1}  d\!\cos{\theta_2} 
\int_0^{2\pi}\frac{d\phi_1}{2\pi} 
\int_0^{2\pi}\frac{d\phi_2}{2\pi}  \\ 
\frac{Q^4}{16}\int_{0}^1 x_2 dx_2
\int_{x_2}^{1} x_1 dx_1\, \Theta
 \left ( \frac{x_2}{2}(1-\cos\theta_2)  - \rho\right) W_2\,,
\end{multline}
whose dominant term at small $\rho$ is $S_2\, \asb^2 \ln^2 1/\rho$.

The integrals over the energy fractions are straightforward. Keeping
only the $\ln^2 1/\rho$ piece and performing the azimuthal average we have
\begin{equation}
S_2 = -
2 C_F C_A 
\int_0^1 d\!\cos\theta_2 \int_{-1}^0 
d\!\cos\theta_1 \,
\Omega_2(\cos\theta_1,\cos\theta_2)\,,
\end{equation}  
where the angular function $\Omega_2$ is 
\begin{equation}
\Omega_2 =  \frac{2}{(\cos\theta_2 -
  \cos\theta_1)(1-\cos\theta_1)(1+\cos\theta_2)}\,.
\end{equation}
Integrating over the polar angles we obtain
\begin{equation}
  S_2 = -\CF \CA \frac{\pi^2}{3}\,.
\end{equation}
We note that possibly related ``$\pi^2$'' terms have been discussed in
\cite{Basics}.

\section{Resummation}
\label{sec:AllOrders}

To perform the all orders resummation of $\cS(\as L)$ we need to
consider general configurations of soft gluons. We define the
probability $P_\cC(L)$ for a given configuration $\cC$ at a resolution
scale $L$ (\ie not resolving gluons with energies less than $Q e^{-L}$).
Given $P_\cC$, the probability $P_{\cC'}$ of a configuration $\cC'$
with one extra gluon with scale $L' > L$, polar angle $\theta'$ and
azimuth $\phi'$ is (see for example \cite{BCM})
\begin{equation}
  \label{eq:BranchProb}
  P_{\cC'}(L') = \asb\!\left(L'\right) \Delta_\cC(L,L')\,
  F_\cC(\theta',\phi')
  P_{\cC} (L) \,,
\end{equation}
where the form factor is given by
\begin{equation}
  \label{eq:Sudakov}
   \ln \Delta_\cC(L,L') = -\int_L^{L'} dL''
   \int d\!\cos\theta \,d\phi \;  \asb\!\left(L''\right)
   F_\cC(\theta,\phi) \,,
\end{equation}
and $F_\cC(\theta,\phi)$ describes the angular and colour-structure of
the radiation pattern from configuration $\cC$. The angular
integrations in \eqref{eq:Sudakov} diverge whenever the angle is
collinear to one of the emitting particles. In practice it is
therefore convenient to introduce an angular cutoff $\epsilon$ to
regulate these divergences both in the real emissions and the virtual
corrections.

We obtain $\cS$ by calculating the probability of there being no
emissions in $\HR$ down to a scale $L$, divided by the corresponding
probability had the only source of emissions been the original
$q\qbar$ pair. This gives
\begin{equation}
  \label{eq:cSAllOrders}
  \cS(\as L) = \frac{1}{\sqrt{\Delta_{ab}(L)}} 
  \sum_{\cC | \HR\,\mathrm{empty}}  P_\cC(L)\,, 
\end{equation}
where the sum runs over all configurations $\cC$ which contain no
emissions in $\HR$.

\subsection{Large-$\boldsymbol{\NC}$ limit and Monte Carlo
  implementation} 

In practice two considerations make it difficult to implement the
above approach analytically. One is that the colour algebra involved
in the determination of $F_\cC$ becomes progressively more complicated
as the number of gluons increases. The other problem is simply that
the treatment of the geometry of the many large-angle gluon ensemble
quickly becomes prohibitive. The first problem can be partially
solved by taking the large-$\NC$ limit. To address the issue of the
geometry we shall use a Monte Carlo approach.

\begin{figure}[tb]
  \begin{center}
    \resizebox{\textwidth}{!}{\input{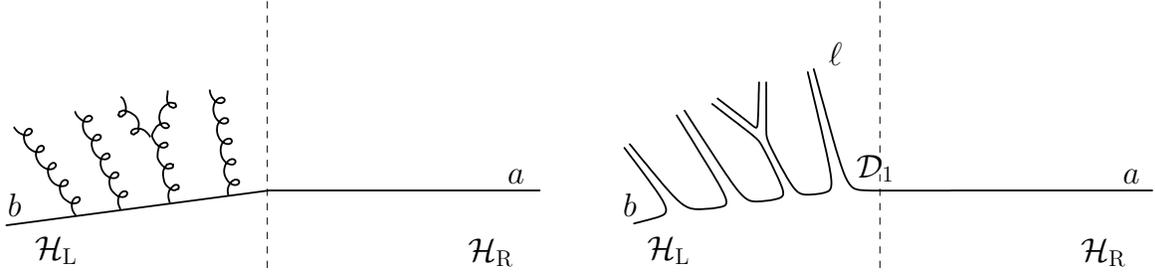}}
    \caption{Left: the kind of diagram which must be considered in
      the calculation of $\cS$. Right: the same diagram represented in
      the large-$\NC$ limit, with gluons shown as pairs of colour
      lines and quarks as single colour lines. }
    \label{fig:largenc}
  \end{center}
\end{figure}

In the large-$\NC$ limit, one can represent gluons by pairs of
colour-anticolour lines, as illustrated in figure~\ref{fig:largenc}.
When squaring the amplitude one ignores contributions that in terms
of their colour flow are topologically non-planar, because they are
suppressed by powers of $1/\NC^2$ \cite{'tHooft:1974jz}.  As a result,
for an ensemble of $n$ gluons, $F_\cC$ just reduces to a sum of
independent emission intensities from $n+1$ independent dipoles:
\begin{equation}
  F_\cC(\theta_k, \phi_k) = \sum_{\mathrm{dipoles}-ij} 
   \frac{2\CA}{(1 - \cos \theta_{ik})(1 - \cos \theta_{kj})}\,.
\end{equation}
When the dipole $ij$ radiates a gluon $k$ it splits into two
dipoles, $ik$ and $kj$. Thus the dipole structure is determined by the
history of the gluon branching.

Such a branching pattern can be very naturally implemented using a
Monte Carlo algorithm. At first sight one might envisage calculating
the two factors in \eqref{eq:cSAllOrders} separately, using the
$F_\cC$ to generate the distribution of radiation for each new
configuration. However because of the collinear divergence along the
direction of the quark in $\HR$, only a tiny fraction of events would
be free of emissions in $\HR$ and so contribute to the sum in
\eqref{eq:cSAllOrders}. The sum would therefore have a large relative
error, which would translate to a large absolute error on $\cS$
because of the division by the small quantity $\sqrt{\Delta_{ab}(L)}$.
  
Instead a more efficient procedure involves moving the division by
$\sqrt{\Delta_{ab}(L)}$ directly into the calculation of the $P_\cC$.
This can be achieved using a modified radiation intensity, $\Ft_\cC$ (for
both the emissions and the virtual corrections),
\begin{equation}
  \Ft_\cC(\theta,\phi) = F_\cC(\theta,\phi) - F_{ab}(\theta,\phi)
  \Theta(\theta)\,,
\end{equation}
where one subtracts out the radiation intensity $F_{ab}$ which would
have been produced by the original $q\qbar$ pair (in the large-$\NC$
limit). One calculates quantities $\Pt_\cC$ using analogs of
eqs.~\eqref{eq:BranchProb} and \eqref{eq:Sudakov} with $F_\cC$
replaced by $\Ft_\cC$ and then $\cS$ is simply given by
\begin{equation}
  \label{eq:cSAllOrdersV2}
  \cS(\as L) = \sum_{\cC | \HR\,\mathrm{empty}}  \Pt_\cC(L)\,.
\end{equation}
It should be kept in mind that since $\Ft_\cC$ is negative in certain
regions of phase space one loses a strict probabilistic interpretation
for the $\Pt_\cC$. Nevertheless the sum over configurations is
well-defined and meaningful.

The exact details of the Monte Carlo algorithm are given in the
appendix. Here we restrict ourselves to giving a parameterisation for
$\cS$ obtained by fitting to the Monte Carlo results:
\begin{equation}
  \label{eq:MainResult}
  \cS(\as L) \simeq \exp\left( - \CF\CA \frac{\pi^2}{3}
     \left( \frac{1 + (at)^2}{1 + (bt)^c}\right)t^2\right),
\end{equation}
with
\begin{equation}
  \label{eq:tdef}
  t (\as L) = \frac{1}{2\pi}\int^1_{e^{-L}} \frac{dx}{x} \as(xQ) = 
  \frac{1}{4\pi \beta_0} \ln \frac{1}{1 - 2\beta_0 \as L}\,,
\end{equation}
where $\beta_0 = (11\CA - 2\nf)/(12\pi)$ and
\begin{equation}
  a = 0.85\CA\,,\qquad b = 0.86\CA\,, \qquad c = 1.33\,.
\end{equation}
The parameterisation should be accurate to the order of a few percent
(better in most of the region) for $t <0.7$, corresponding to $1 -
2\as\beta_0L \gtrsim 0.005$.\footnote{The accessible range of $t$ is limited
by two issues: firstly only a small fraction of events are generated
at large $t$, requiring considerable statistics in order to
investigate that region; and secondly because an accurate
determination of $\cS$ at large $t$ requires a very small angular
cutoff, which leads to there being many dipoles in an event, and a
consequent slowing down of the evolution.}

Actually, for the purposes of the fit one replaces $\CF\CA$ in
\eqref{eq:MainResult} with $\CA^2/2$ since the Monte Carlo works in
the large-$\NC$ limit. But for use in phenomenology one wishes to have
the exact colour structure at least at $\order{\as^2}$, hence the use
of $\CF\CA$ in \eqref{eq:MainResult}.

\section{Checks and conclusions}
\label{sec:Event2}

It is useful to check our results against fixed order results from the
next-to-leading order Monte Carlo program Event2 \cite{CSDipole}.
First it is necessary to determine the constant terms $C_1^{(q)}$ and
$C_1^{(g)}$, which are obtained by requiring consistency between
\eqref{eq:finalform} and a full $\order{\as}$ calculation.  It is
straightforward to show that they are given by
\begin{equation}
  \label{eq:rhoC1}
  C_1^{(q)} = \frac{1}{2} 
         \left(C_1^{\tau} - r_3\right)\,, \qquad 
  C_1^{(g)} = \frac{r_3}{2}\,,
\end{equation}
where $C_1^{\tau}$ is the constant determined for the thrust in \cite{CTTW}
and $\asb r_3$ is the probability of there being a hard gluon along
the thrust axis, as given in \cite{rhoLight}:\footnote{Only in  the hep-ph
  arXiv version 1.}
\begin{equation}
  \label{eq:r3}
  r_3 = \CF \left( 2\ln^2 2 - \frac54 \ln3 +
    4\mathrm{Li}_2\!\left(-\frac12\right) + \frac{\pi^2}{3} -
    \frac16\right)
  \simeq 0.917 \CF\,.
\end{equation}

\begin{figure}[tb]
  \begin{center}
    \epsfig{file=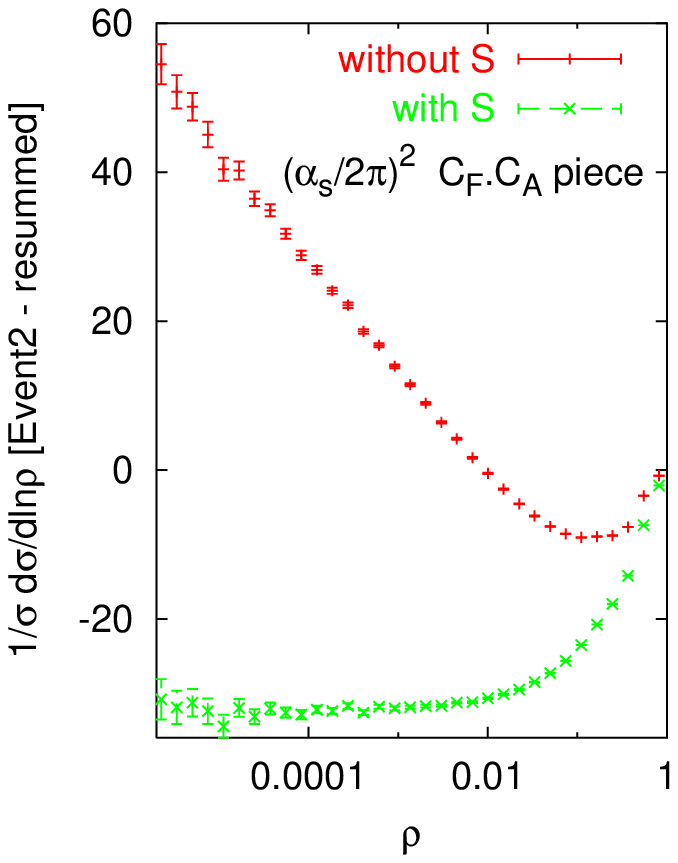}\;\;\;\;
    \epsfig{file=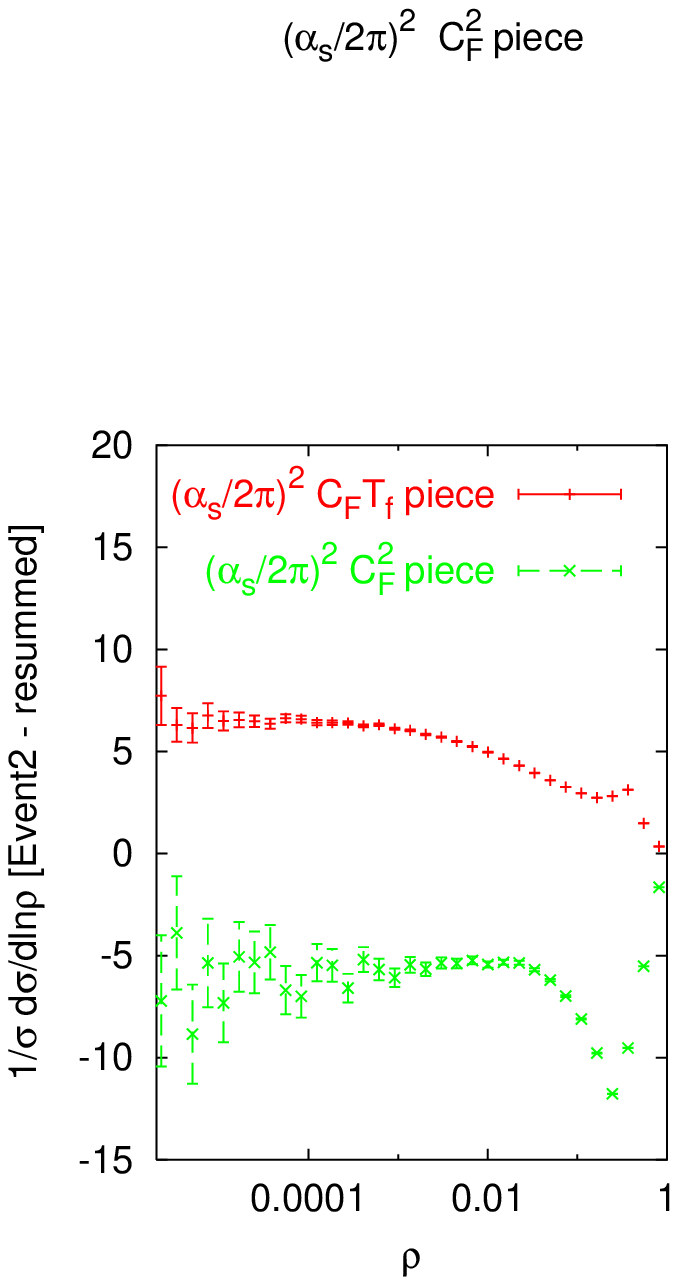}
    \caption{The difference between  fixed order results from Event2
      and the expansion of the resummed results. The left-hand plot
      shows results for the coefficient of $\asb^2\CF\CA$ --- the two
      sets of points correspond to neglecting or accounting for $\cS$.
      The right-hand plot shows the coefficients of the $\CF T_f$ and
      $\CF^2$ pieces.}
    \label{fig:event}
  \end{center}
\end{figure}

The plots in figure~\ref{fig:event} show the difference between the
$\order{\as^2}$ exact and resummed distributions. To NLL accuracy we
aim to account for terms down to $\as^2 L^2$ in $\Sigma$ and therefore
$\as^2 L$ in the distribution. Accordingly the difference should be
independent of $\rho$ for small $\rho$. One can see that this is the
case for each individual colour factor. We also show that without the
contribution $\cS$ calculated in this paper the $\CF\CA$ part of the
result is not consistent with Event2.


Using eq.~\eqref{eq:heavylight} it is possible to combine our results
for the single-hemisphere mass together with those of \cite{CTTW} for
the heavy-jet mass in order to obtain predictions correct to
single-log level for the light-jet mass (including the appropriate $C_1$
terms).

We point out that the form computed for $\cS$ applies to all
single-hemisphere 2-jet event-shapes that for large-angle gluons have
a value of the order of the transverse momentum: in $\ee$ this means
the hemisphere broadening (for the constant terms a relation analogous
to \eqref{eq:rhoC1} applies, with $C_1^{\tau}$ replaced with
$C_1^{B_T}$). In DIS our form for $\cS$ applies to the current
hemisphere jet-mass, the $C$-parameter, and the thrust and broadening with
respect to the thrust axis. The detailed phenomenology of the DIS
variables will be considered elsewhere \cite{DSDIS}. It indicates that
the inclusion of $\cS$ reduces the peak height (in $d\sigma/d\rho$) by
about $30\%$ before 
matching to the $\order{\as^2}$ calculation, and by about $20\%$ after
matching (using an $R$-type matching \cite{CTTW}).

For the narrow-jet $K_{out}$ and jet masses in multi-jet events
(including hadron-hadron event shapes) different forms will apply
because of the more complicated geometry of the underlying hard event,
but the Monte Carlo algorithm described here can be adapted to those
cases as well.

\par \vskip 1ex
\noindent{\large\bf Acknowledgments}

We wish to thank Stefano Catani for helpful discussions and Andrea
Banfi, Yuri Dokshitzer, Pino Marchesini and Giulia Zanderighi for
helpful comments and suggestions.  One of us (GPS) would like to
acknowledge the hospitality of the DESY theory group while part of
this work was carried out. We are also grateful to the INFN, Sezione
di Milano and the Universit\`a di Milano Bicocca for the use of
computing facilities.

\appendix
\section*{Appendix}

Below we give details of the Monte Carlo algorithm used to calculate
  $\cS$.\footnote{We 
  present a version of the algorithm which has weighted events; there
  exists also a slightly more efficient version of the algorithm with
  unweighted events, but it requires some integrations to be done
  analytically and is therefore less easily generalisable to
  observables with more complex boundaries in phase space such as jet
  masses in multi-jet events.}

For each dipole $\cD_i$, the collinear angular cutoff $\epsilon$ introduced
in section~\ref{sec:AllOrders} defines a rapidity interval $\Delta\eta_i$
for emission in the dipole c.o.m.\ frame. We shall denote by $\cD_1$
the dipole which has as one of its sources the quark in $\HR$. 
The evolution variable used will be $t$, defined in
\eqref{eq:tdef}. This will account for the running of the coupling.

One starts an event with a quark-antiquark dipole (labelled $ab$), $t=0$
and a weight $w = 1$. One then goes through the following steps to
determine an event's contribution to the quantity $-d\cS/dt$:
\begin{itemize}
\item[1.] Establish $\Delta\eta_\tot = \sum_i \Delta\eta_i$. Increase
  $t$ by an amount $\Delta t$ chosen according to a distribution
  proportional to
  \begin{equation}
    \exp\left( -2\CA \Delta\eta_\tot \Delta t\right)\,.
  \end{equation}
\item[2.] Choose a dipole at random, such that the probability of
  obtaining $\cD_i$ is $\Delta\eta_i/\Delta\eta_\tot$. Create a gluon
  uniformally in rapidity and azimuth in the dipole centre of mass
  frame, such that in the lab frame its angle relative to either of
  the dipole sources is larger than $\epsilon$. 
\item[3.] There are now various possibilities:
  \begin{itemize}
  \item The gluon is in $\HL$: insert it into dipole $i$ so as to
    create two new dipoles (removing the old one from the list). Go to
    step 1.
  \item The gluon is in $\HR$ and was not emitted from $\cD_1$: add
    $w$ to the bin corresponding to the value of $t$ and start a new
    event.
  \item The gluon is in $\HR$ and was emitted from $\cD_1$: denote the
    momenta of its sources by $p_\ell$ and $p_a$ (the latter is just the
    $\HR$ quark momentum) and the new gluon momentum by $p_g$. One
    determines the value of 
    \begin{equation*}
      X = 1 - \frac{F_{ab}}{F_{a\ell}} 
         =  1 - \frac{(ab)}{(gb)} \frac{(g\ell)}{(a\ell)}\,.
    \end{equation*}
    If $X>0$ then with probability $X$ we add $w$ to current $t$-bin
    and start a new event; with probability $1-X$ we throw away the
    gluon that has just been generated and go to step 1. If $X<0$ then we
    add $Xw$ to the current $t$ bin, multiply $w$ by $1-X$, and go to
    step 1. This procedure is a simple extension to negative
    probabilities of the veto algorithm and corresponds to replacing
    $F_\cC$, used implicitly in the previous steps, with $\Ft_\cC$.
  \end{itemize}
\end{itemize}


\end{document}